\documentclass[final,preprint,draft,showpacs,showkeys,tightenlines,aps,superscriptaddress,fleqn]{revtex4}% 
\usepackage{graphicx}
\usepackage{dcolumn}
\begin{document}
\preprint{PNU-NTG-12/2004}
\preprint{PNU-NuRI-09/2004}

\title{Nonlinear Dynamical Behaviors of the Equilibrium Longitudinal
Distribution by Localized Wakes in an Electron Storage Ring} 
\author{S.-H. Park\footnote{Present address: LG.PHILIPS LCD,
161, Imsoo-dong, Kumi, Kyungbuk 730-350, KOREA}}
\email{shpark@lgphilips-lcd.com}
\affiliation{Department of Physics \& Nuclear Physics and Radiation
Technology Institute (NuRI), Pusan National University, Busan 609-735, Korea}
\author{H.-Ch. Kim}
\email{hchkim@pusan.ac.kr}
\affiliation{Department of Physics \& Nuclear Physics and Radiation
Technology Institute (NuRI), Pusan National University, Busan 609-735, Korea}
\author{J.K. Ahn}
\email{ahnjk@pusan.ac.kr}
\affiliation{Department of Physics \& Nuclear Physics and Radiation
Technology Institute (NuRI), Pusan National University, Busan 609-735, Korea}
\author{E.-S. Kim }
\email{eskim1@postech.ac.kr}
\affiliation{Pohang Accelerator Laboratory, POSTECH, San 31, Pohang,
Kyungbuk 790-784, Korea}

\begin{abstract}
We investigate the effects of a localized constant wake and 
a localized linear wake on the longitudinal beam distribution in equilibrium.  
We also examine the effects of the
composite wake on the particle distribution when the constant wake and
linear wake simultaneously exist in the ring. While moving around the 
parameter space, the system can show bifurcation phenomena and
transition features between periodic states.  We study the dynamical 
states in a beam, using the multi-particle tracking method in a
Gaussian approximation.  They show good qualitative agreement. 
\end{abstract} 
\pacs{29.20.Dh}
\keywords{wakes, bifurcation, electron storage ring, multi-particle
tracking}  

\maketitle

\section{Introduction}
When a beam is stored in a storage ring, the beam is affected by
a wake force that is generated by an electromagnetic interaction
between the beam and the components of the ring. The wake force 
disturbs the distribution of particles in the beam.  Various
approaches to investigate the particle distribution have been
performed.  It was usually assumed that the wake force was
averaged over one turn and was distributed uniformly in the
ring.  In fact, however, the wake sources may be localized. Therefore,
such assumptions of an averaged and uniform wake force may
cause different dynamical behaviors of the system from real
ones in many cases.  To see the effects due to such position-dependent
wake forces, Hirata~\cite{hirata1} investigated the particle distribution
in a beam, assuming a constant localized
wake.  Hirata $et$ $al$.~\cite{hirata2} then
showed that the equilibrium bunch length in electron storage rings
could exhibit a cusp-catastrophe behavior.  Kim {\em et
al.}~\cite{eskim} found that several multi-periodic states in 
particle distributions of a beam could also exist for a constant
wake. 

The aim of the present work is at investigating dynamic behaviors of 
the longitudinal beam distribution in the presence of the wake that
has two components localized at one position of the ring. Here,
we consider the linear wake as well as the constant one as
sources of the localized wakes.  It is also of great interest to
see how the longitudinal beam distribution behaves dynamically under the
composite constant and linear wakes in a ring.  The dynamical
behaviors of the beam distribution will be examined with the damping time,
the strength of the wake force, and the synchrotron tune considered.
It will be shown that the dynamic system under two localized wake
sources may present quite different dynamic behavior from that with an
individual wake source taken into account. 

The paper is organized as follows: In Section 2, we briefly describe
the model which is used in order to investigate the
equilibrium logitudinal distribution of the beam particles in a Gaussian
approximation.  In Section 3, we study the dynamical states of the
distribution of particles in a beam.  To check the validity of the
results from the present model, we also compare them with those from
the multi-particle tracking in Section 4.  Finally,   
Section 5 is devoted to discussion and conclusion. 

%%%%%%%%%%%%%%%%%%%
\section{The Model}
%%%%%%%%%%%%%%%%%%%

\subsection{Longitudinal dynamics}
We assume  that there are two localized wake sources at one
position of the ring.  To describe dynamics of the particle
distribution in longitudinal phase space, we introduce the following
normalized longitudinal variables:
\[
  x_1 = \frac {\rm{longitudinal\ displacement}}{\sigma_z}, \
  x_2 =\frac{\rm{energy\ deviation}}{\sigma_E} ,
\]
where $\sigma_z$ is the natural bunch length and $\sigma_E$ stands for
the natural energy spread.  The center of the bunch is located at
$x_1$=0 such that $x_1 > 0$ corresponds to the rear part of the 
bunch.  The motion of a particle in a ring can then be modeled as follows: \\

1) Radiation

\begin{eqnarray}
\left(\begin{array}{c}
                  x_1^{\prime} =x_1    \\
                  x_2^{\prime} = \Lambda x_2 + (1 - \Lambda^2)^{1/2} \hat{r}
              \end{array}    \right )
\end{eqnarray}

2) Wake

\begin{eqnarray}
\left(\begin{array}{c}
                  x_1^{\prime} =x_1    \\
                  x_2^{\prime} = x_2 - \phi(x_1)
              \end{array}    \right )
\end{eqnarray}

3) Synchrotron oscillation

\begin{eqnarray}
       \left( \begin{array}{c}
                x_1^{\prime}    \\
                x_2^{\prime}
             \end{array}        \right) =  U \left( \begin{array}{c}
                   x_1    \\
                x_2
             \end{array}         \right)   ,
    \label{syno}
\end{eqnarray}
where
\begin{eqnarray}
U =
        \left( \begin{array}{cc}
    \cos(2\pi\nu)  & \sin(2\pi\nu)  \\
    -\sin(2\pi\nu) & \cos(2\pi\nu)
             \end{array}         \right)  .
    \label{U}
\end{eqnarray}
In the above equations,  $\Lambda =\exp(-2/T_e)$, $T_e$ being the
synchrotron damping time divided by the revolution time, 
$\nu$ the synchrotron tune and $\hat{r}$ a Gaussian random variable
with zero mean and unit standard deviation.

After one turn in the ring, the motion of a particle is described
by
\begin{eqnarray}
       \left( \begin{array}{c}
                x_1^{\prime}    \\
                x_2^{\prime}
             \end{array}        \right) = U \left( \begin{array}{c}
   x_1 \\
   \Lambda x_2 + (1- \Lambda^2)^{1/2}\hat{r} - \phi(x_1)
             \end{array} \right).
    \label{syno1}
\end{eqnarray}
 The wake force is given by
\begin{eqnarray}
   \phi(x_1) = \int_{0}^{\infty} \rho(x_1-u)W(u)du,
\end{eqnarray}
where $\rho(x)$ is the longitudinal charge density normalized to
unity, and $W(u)$ stands for the longitudinal wake function 
multiplied by $eQ/\sigma_E$, where $e$ is the electron charge and $Q$
the total charge in a bunch. 

In the present work, we consider the constant as well as the linear
wake functions: $W(u) = a \Theta(u) $ ($\Theta$ being the unit step
function) and $W(u) = b u$, where $a$ and $b$ denote the strengths of
the wake. 

\subsection{ Gaussian model}
We assume the wake to disappear at a short distance behind the
particles producing it so that one can neglect multi-turn 
effects of the wake. Generally, leading particles in a bunch lose 
energy due to wake fields.  In order to meet this condition, we
note that the signs of $a$ and $b$ in the wake functions should be
positive.

Since it is not realistic to observe individual particles, we are more
interested in statistical quantities such as
\begin{eqnarray}
\bar{x}_i = \langle x_i \rangle,\  \sigma_{ij} = \langle
(x_i-\bar{x}_i) (x_j - \bar{x}_j) \rangle, 
\end{eqnarray}
where $i$, $j$ are either $1$ or $2$, which are the moments of the
phase-space distribution $\Psi(x_1,x_2)$. In reality, we require all
of higher order moments to reproduce $\Psi(x_1,x_2)$.  We always
approximate $\Psi(x_1,x_2)$ as 
\begin{eqnarray}
\Psi(x_1,x_2) = \frac{1}{2\pi\sqrt{{\rm det}\sigma}} \times \exp[-\frac{1}{2}
\sum_{i,j}\sigma^{-1}_{i,j}(x_i-\bar{x}_i) (x_j - \bar{x}_j)].
\end{eqnarray}
For simplicity, however, we assume here that the distribution function 
in phase space always remains Gaussian even under the influence of the
wake force.  We thus need to consider only the first- and second-order
moments. 

With the same treatment as in Ref.~\cite{hirata1} used, each mapping
in Eqs.~(1)-(3) can be given as :
\\
1) Radiation
     \begin{eqnarray}
     \bar{x}_1^{\prime} &=& \bar{x}_1,\  \bar{x}_2^{\prime} =\Lambda \bar{x}_2, \\
\sigma_{11}^{\prime}&=&\sigma_{11},\ \sigma_{12}^{\prime}=\Lambda
\sigma_{12}, \nonumber \\ 
   \sigma_{22}^{\prime} &=& \Lambda^2 \sigma_{22}+(1-\Lambda^2)
\end{eqnarray}
2) Wake
   \begin{eqnarray}
     \bar{x}_1^{\prime} &=&\bar{x}_1, \
     \bar{x}_2^{\prime} = \bar{x}_2 - \langle \phi \rangle, \\
     \sigma_{11}^{\prime} &=& \sigma_{11},\
     \sigma_{12}^{\prime} = \sigma_{12} - \langle (x_1 - \bar{x}_1 ) \phi)
     \rangle, \nonumber \\ 
     \sigma_{22}^{\prime} &=&\sigma_{22}-2 \langle(x_2-\bar{x}_2)\phi\rangle
    +\langle\phi^2\rangle-\langle\phi\rangle^2. \nonumber \\
    \end{eqnarray}
Further calculations lead us to obtain for the constant wake~\cite{hirata2}
   \begin{eqnarray}
    \bar{x}_1^{\prime} &=& \bar{x}_1,\    \bar{x}_2^{\prime} =\bar{x}_2 -a/2, \\
     \sigma_{11}^{\prime} &=& \sigma_{11},\
     \sigma_{12}^{\prime} =
     \sigma_{12}-a\frac{\sqrt{\sigma_{11}}}{2\sqrt{\pi}}, \nonumber \\ 
  \sigma_{22}^{\prime} &=&
     \sigma_{22}-a\frac{\sqrt{\sigma_{12}}}{\sqrt{\pi\sigma_{11}}}+a^2/12. 
 \end{eqnarray}

As for the linear wake, we derived the following relations from 
   eqs. (11) and (12)
   \begin{eqnarray}
    \bar{x}_1^{\prime} &=& \bar{x}_1,\
    \bar{x}_2^{\prime} =\bar{x}_2 -b \sqrt{\frac{\sigma_{11}}{\pi }}, \\
     \sigma_{11}^{\prime} &=& \sigma_{11},\
     \sigma_{12}^{\prime} = \sigma_{12}-\frac{b\sigma_{11}}{2}, \nonumber \\
  \sigma_{22}^{\prime} &=& \sigma_{22} -b\sigma_{12}
    +b^2\sigma_{11}(0.608998+\frac{1}{\pi}). 
\end{eqnarray}
3) Synchrotron oscillation
       \begin{eqnarray}
              \bar{x}_i^{\prime} &=& \sum_{j}U_{ij}\bar{x}_j , \\
          \sigma_{ij}^{\prime} &=& \sum_{h,k=1}^{2} U_{ij}\sigma_{hk}U_{jk} .
     \end{eqnarray}
Note that the mappings for $\bar{x}_{i}$ and $\sigma_{ij}$ are related
to each other for the constant wake function.  However,  
those for $\bar{x}_{i}$ depend on $\sigma_{11}$ for the linear
wake function due to the fact that we have utilized the
approximation of the Gaussian distribution .

%%%%%%%%%%%%%%%%%%%%%%%%%%%%%%%%%%%%%%%%%%%%%%%%%%
\section{Dynamical Behavior of the Gaussian Model}
%%%%%%%%%%%%%%%%%%%%%%%%%%%%%%%%%%%%%%%%%%%%%%%%%%
If $S$ is the mapping described by Eq.~(14) or Eq.~(16), then a
period-one fixed point $\sigma=(\sigma_{11},\sigma_{12},\sigma_{22})$
is defined by 
\[ \sigma = S(\sigma),   \]
while the period-two fixed point is defined by 
\[ \sigma =S[S(\sigma)]   \]
with $\sigma_1$ and $\sigma_2$ occurring in alternating pairs
at successive turns as
\[ \sigma_{1,2} =S(\sigma_{2,1}). \]
One can easily extend the definition to period-n states.

First, let us consider the dynamical behavior of the system for the
constant wake function.  In a previous work~\cite{eskim}, one of the
authors studied the dynamic states in the parameter space set: $0.01
\leq \nu \leq 0.3$, $1 \leq T_e \leq 1500 $ and $ 0 < a \leq 45$.  The
results are summarized in Table~I, which shows four different 
behaviors, depending on the synchrotron tune.   Those numbers in
Table~I denote existing period states, i.e. 1 stands for the
period-one state, 2 does the period-two state, and 1-2 denotes the
period-one and period-two states existing simultaneously.  The
parameter space shows that several types of equilibrium states such as
period-one, period-two, period-three, period-four can exist stably
depending on the parameter space ($T_e$ and $a$).

In the case of the linear wake function, it is shown that the
period-one state only exists as an equilibrium state.  The tracking of
the mapping for the period-one state is drawn in Fig.~1, where the
features of the trajectories are indicated near the period-one  
fixed point. In the left panel of Fig.~1 the fixed point is depicted
at $(\sigma_{11}, \sigma_{22})=$ ($0.53105$, $0.53105$), whereas in the 
right panel the fixed point is drawn at $(\sigma_{11}, 
\sigma_{22})=$ ($0.05341$, $0.22875$).

When it comes to the composite wake of the constant and linear ones
$(a\Theta + bu)$, Fig. 2 depicts the trackings of the mapping with
the different $b$ for $\nu = 0.20$.  The system plotted in Fig. 2(a)
presents the period-two state, in which the fixed points are located at
$(\sigma_{11}, \sigma_{22}) =$ ($15.43191$, $17.01573$) and ($24.96073$,
$31.95213$).  However, when the strength of the linear wake becomes
larger, the number of the fixed points is reduced to one, as shown in
Fig. 2(b).  Thus the system shows the period-one state that has a
fixed point at $(\sigma_{11}, \sigma_{22}) =$ ($6.31035$, $9.77105$).

\begin{figure}[tb]
\begin{center}
\vspace{0cm}
\includegraphics[width=55mm]{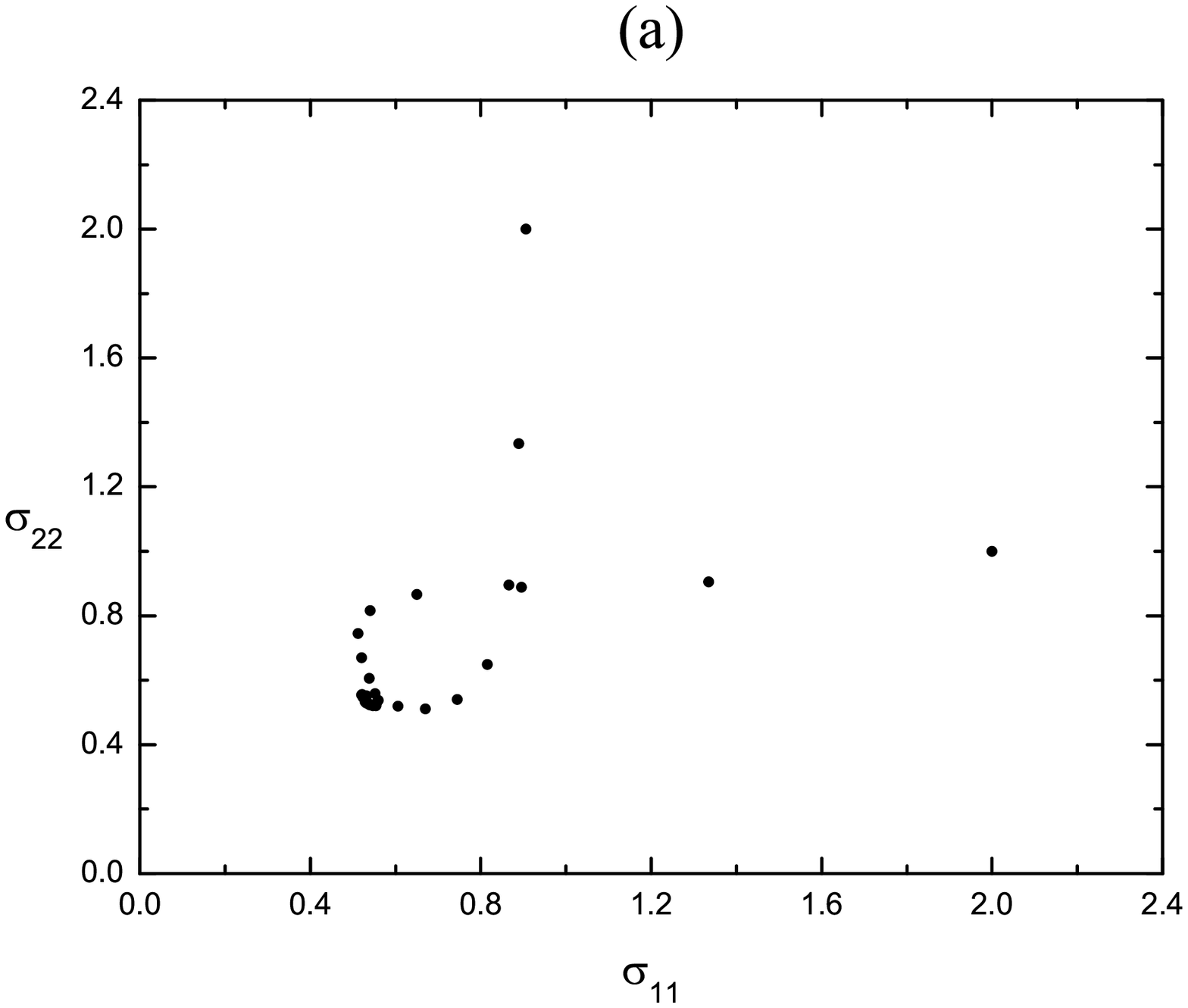}
\includegraphics[width=55mm]{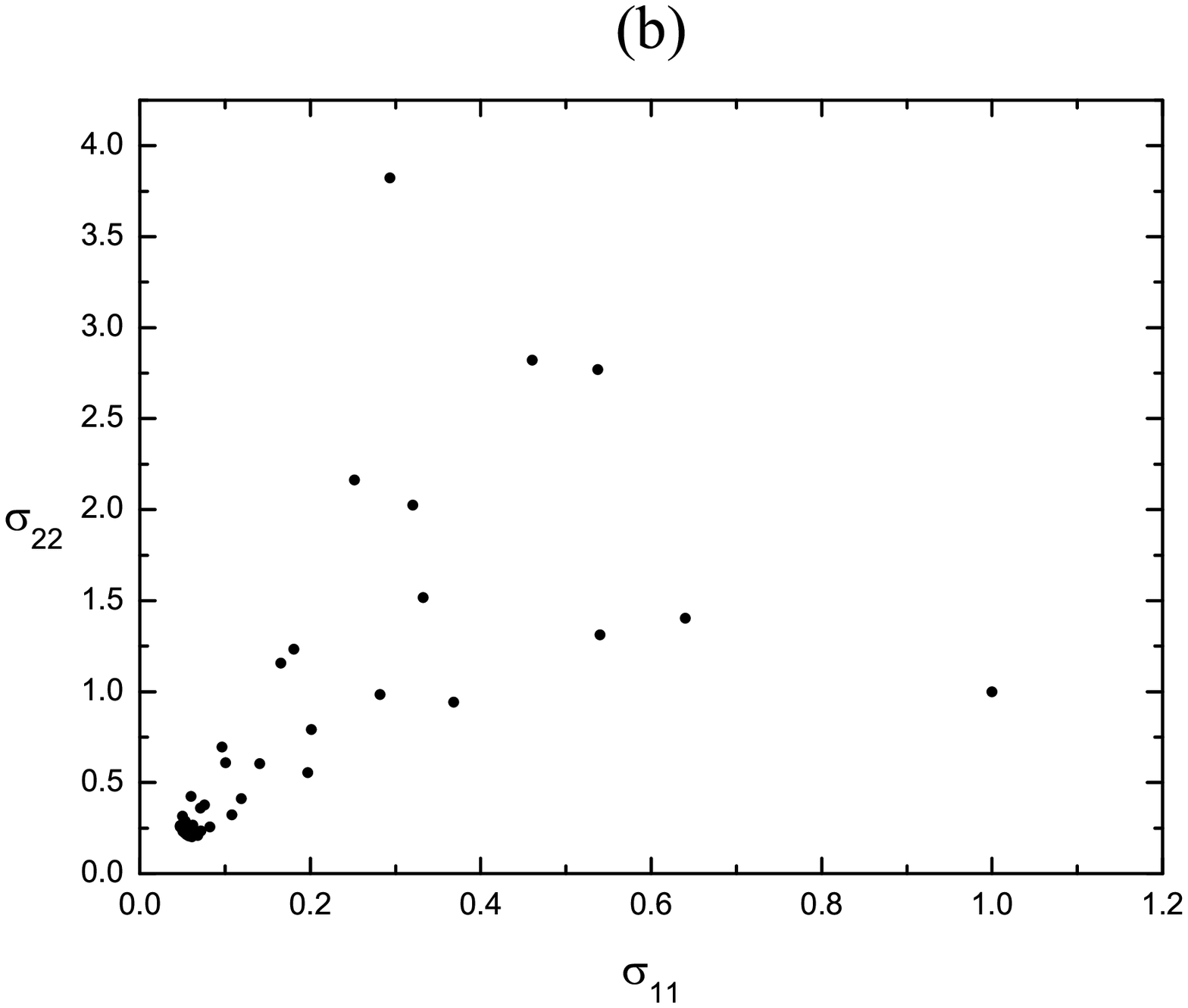}
\end{center}
\caption{Features of trajectories for the linear
wake. The parameters are initially given for the left panel (a) $T_e
$= 10, $b$ = 0.5, $\nu$ = 0.25, $\sigma_{11}$ = 2, $\sigma_{12}$ =
0, $\sigma_{22}$ = 1, while for the right panel (b) $T_e$ = 15, $b$ =
2.0, $\nu$ = 0.05, $\sigma_{11}$ = $\sigma_{22}$ = 1,
$\sigma_{12}$=0.} \label{fig1} 
\end{figure}

\begin{figure}[tb]
\begin{center}
\vspace{0cm}
\includegraphics[width=55mm]{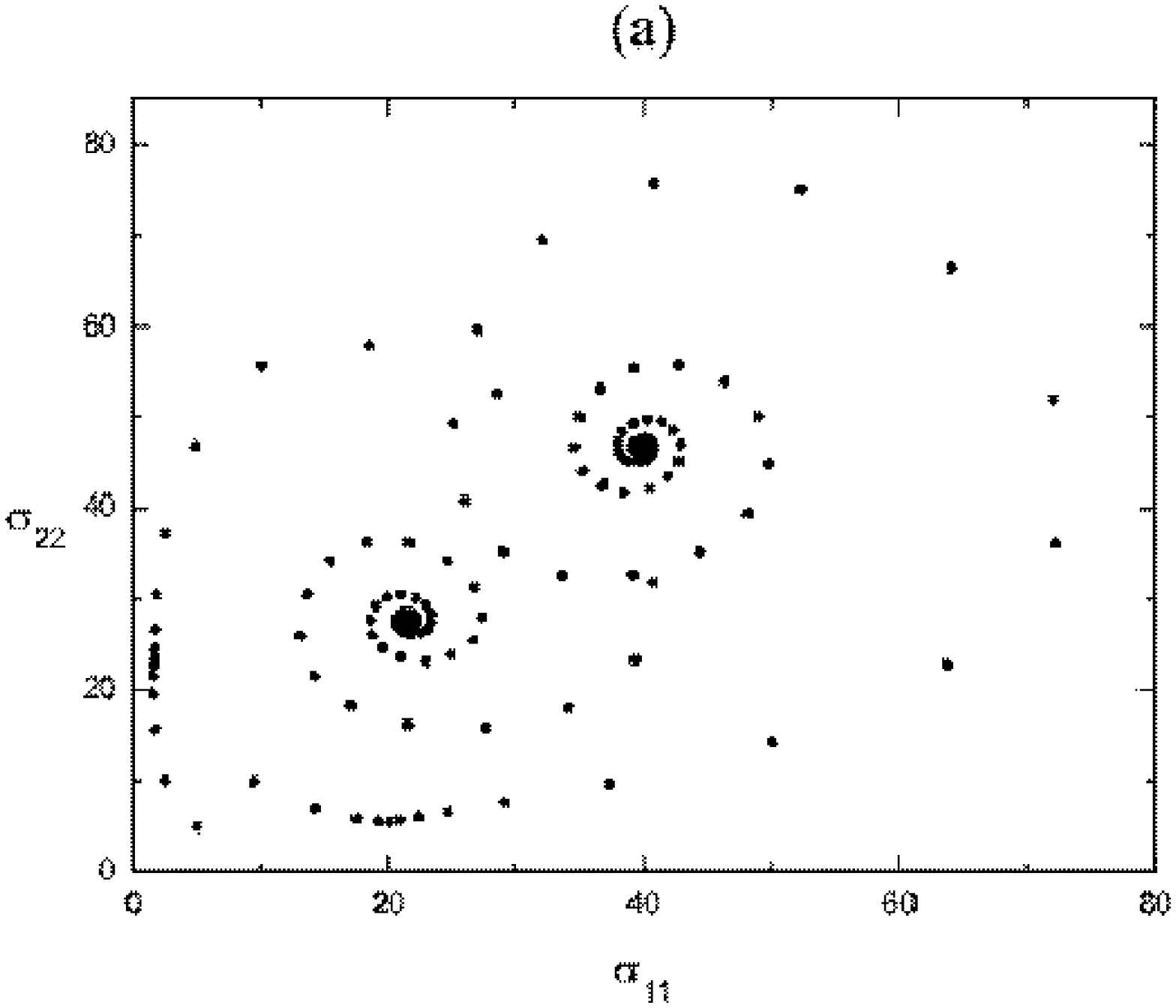}
 \includegraphics[width=55mm]{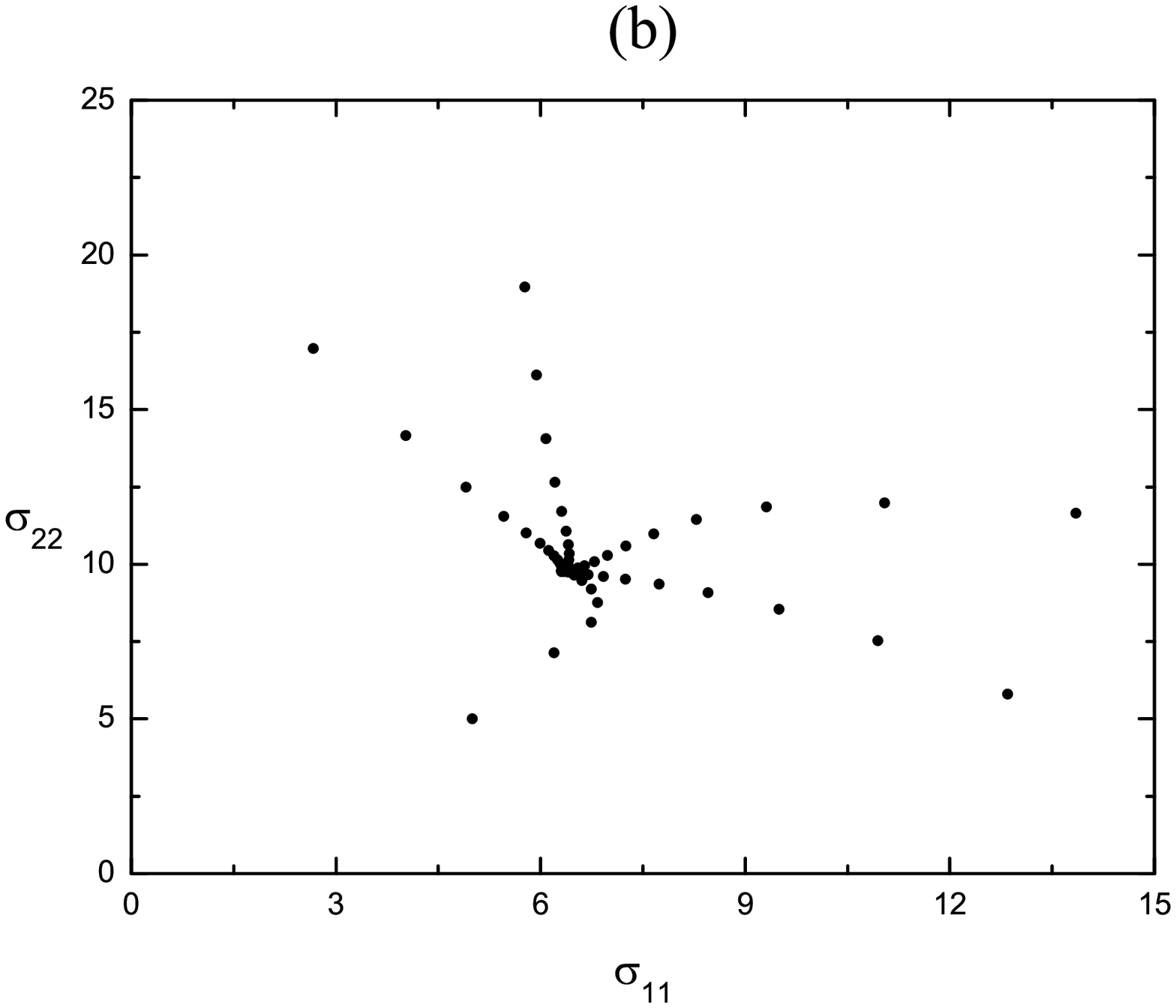}
\end{center}
 \caption{Features of trajectories for the composite
wake.   The parameters are initially given as $T_e$ = 25, $\nu$
= 0.20, $a$ = 10, $\sigma_{11}$ = $\sigma_{22}$ = 5, $\sigma_{12}$
= 0.  $b = 0.25$ for the left panel (a), while $b= 1.0$ for the right
panel (b).} \label{fig1} 
\end{figure}

\begin{figure}[tb]
\begin{center}
\vspace{0cm}
\includegraphics[width=55mm]{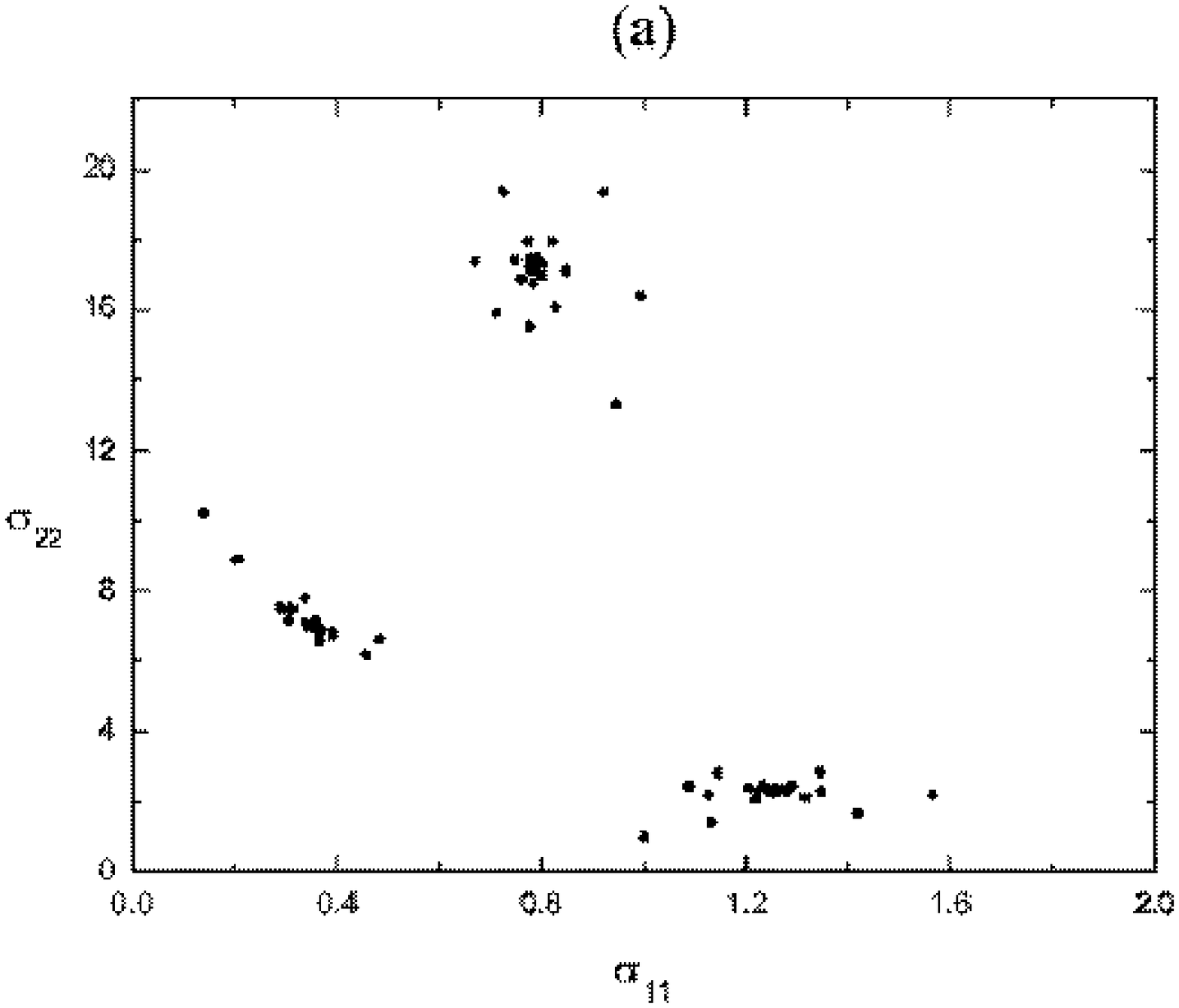}
 \includegraphics[width=55mm]{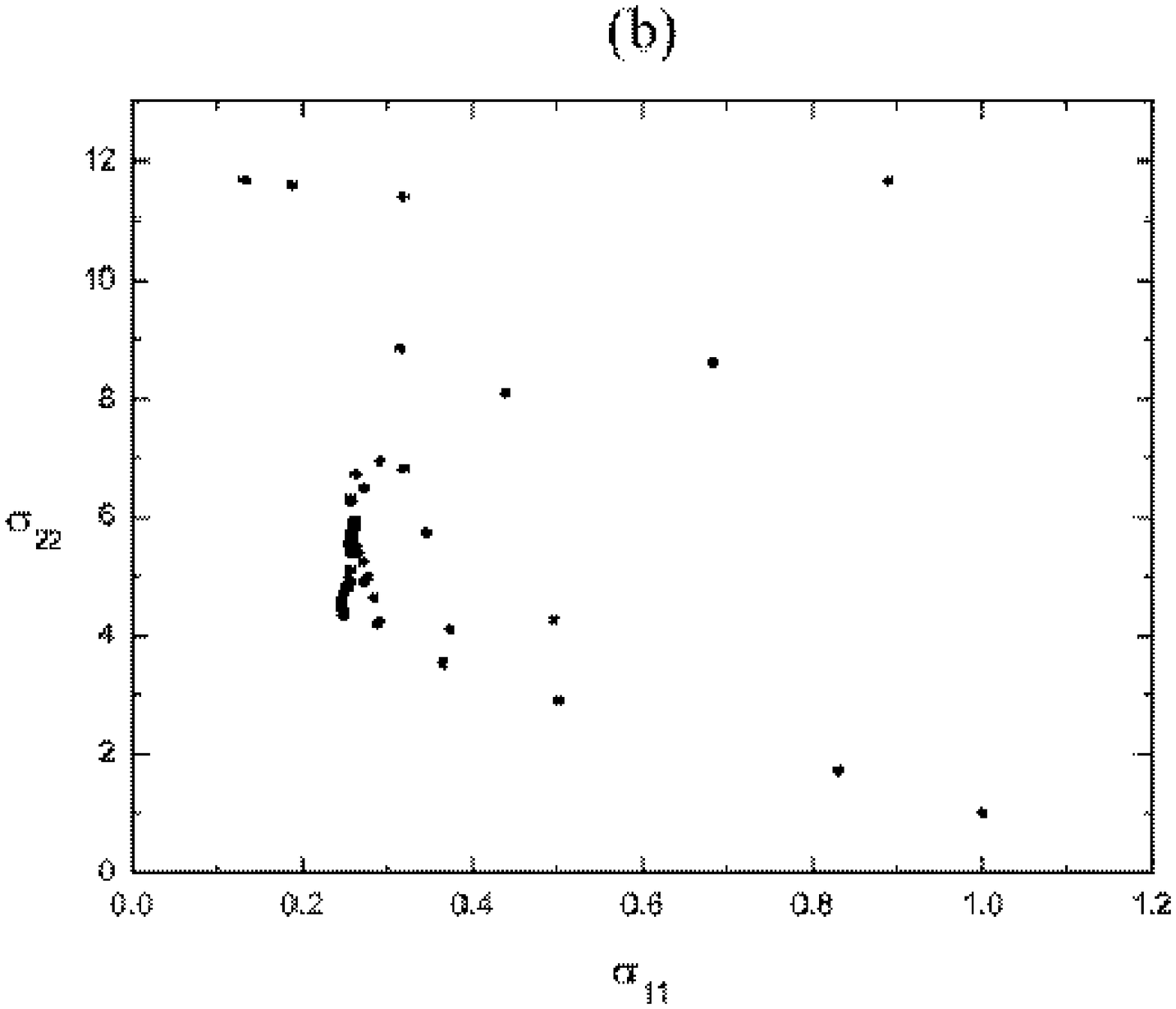}
\includegraphics[width=55mm]{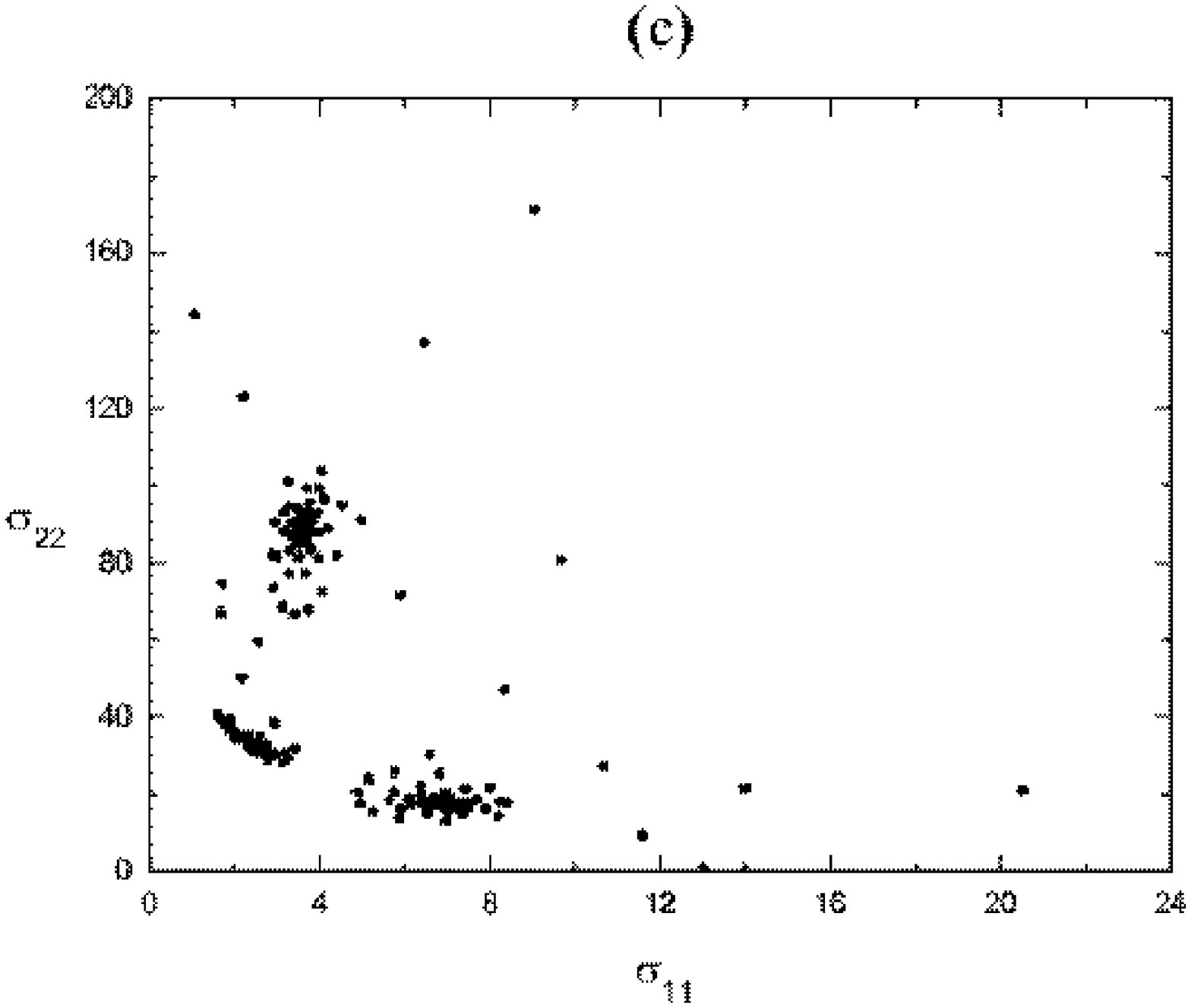}
 \includegraphics[width=55mm]{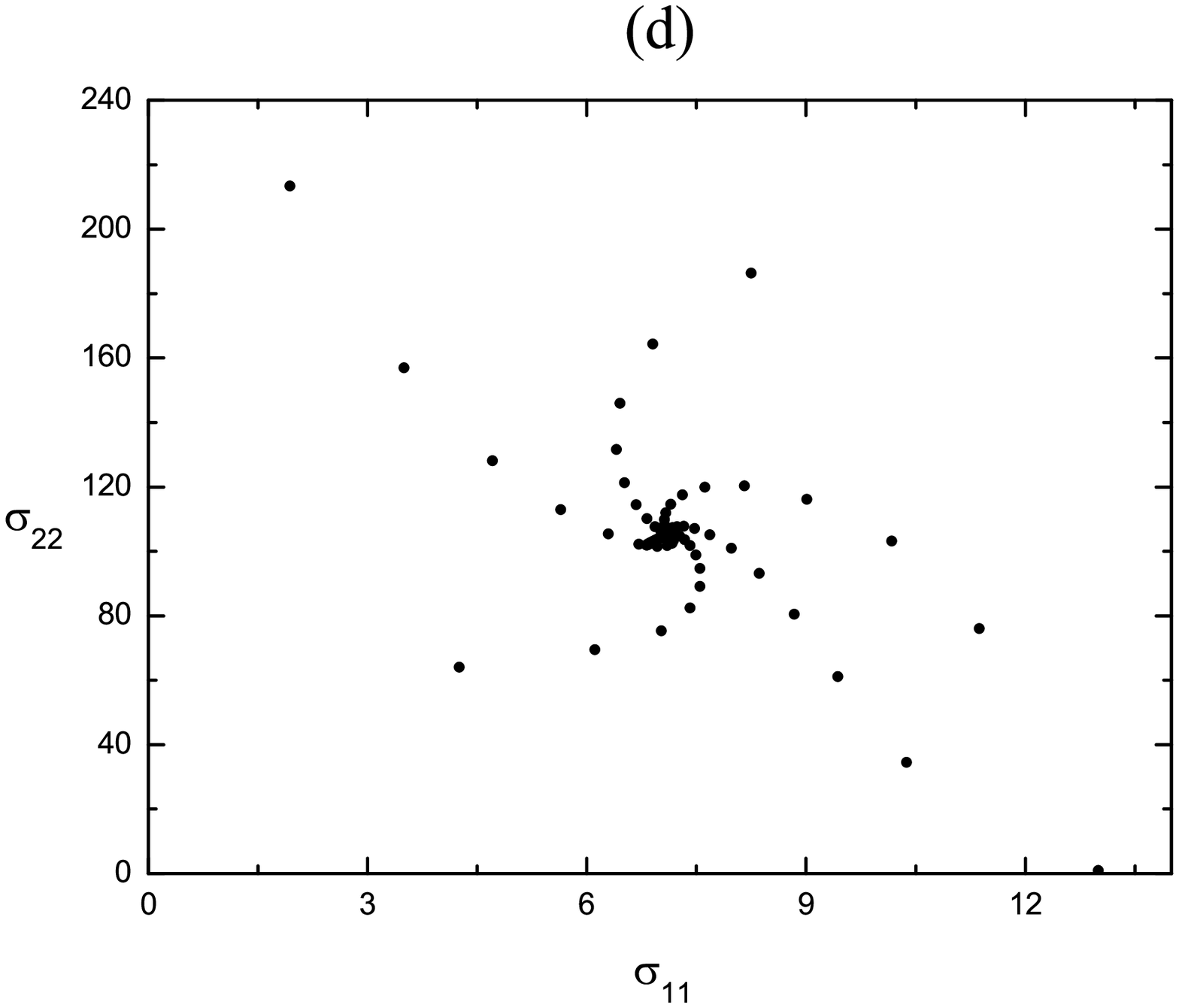}
\end{center}
 \caption{Features of trajectories for the composite
wake.  The parameters are initially given as $T_e$ = 15, $\nu$ = 0.05,
$\sigma_{11}$ = $\sigma_{22}$ = 1, $\sigma_{12}$ = 0,   $a$ = 10.  In
the panel (a) $b= 0.01$, while in (b), (c), and (d), $b$=0.5, 5.0,
and 7.5, respectively.}
\label{fig3}
\end{figure}
Figure 3 shows the trackings of the mapping for the composite wake in
$\nu = 0.05$.  Given $a \neq 0$, $b = 0$, which means that the system
is under the constant wake, the system turns out to be in the
period-three state.  However, when $a \neq 0$ and $b \neq 0$, the
system may exist in different states.  In the panel (a) of Fig. 3 it
is still in the period-three state with $b$ = 0.01. In that case, the
strength of the linear wake is small so that the system shows similar
behavior to the constant wake case. The fixed points are located at
$(\sigma_{11}, \sigma_{22}) =$ ($0.78431$, $17.25226$), ($1.257$,
$2.34851$) and ($0.35848$, $6.9652$).  As the strength of the linear
wake becomes larger, the state of the system starts to get changed.
When $b$ is 0.5, the system is found in the period-two state with
fixed points at $(\sigma_{11}, \sigma_{22}) =$ ($0.26121$, $5.89856$)
and ($0.24996$, $4.39433$).  If $b$ is set to be $5.0$, the system
turns back to the period-three state again.  The fixed points exist at
$(\sigma_{11}, \sigma_{22}) =$ ($6.86404$, $17.82114$), ($2.46592$,
$32.71493$) and ($3.61459$, $89.64147$).  Putting $b=7.5$, we find
that the system lies in the period-one state.  In the panels (a) and
(b) of Fig. 4 and in (c),(d) show the equilibrium of $\sigma_{11}$ for
the constant and the composite wakes with $T_e$ slowly changed,
respectively, while in the panels (a) and (c) the equilibrium of
$\sigma_{11}$ is shown when $T_e$ increases from $0$ to
$30$. In the panels (b) and (d) of Fig. 4, we find the processes
reversed.  There are two transition points in Fig. 4(a): The system
shows the transitions from period-one to period-two and then from
period-two to period-one.  The reverse process for the constant wake
is shown in the panel (b) of Fig. 4 and it also has two transition
points but the transition states are different from those in the panel
(a): It is found that the period-three state changes to the period-two
one and vice versa.  On the other hand, In the panel (c) of Fig. 4  
the composite wake plays a role of changing the period-one to the
period-two state and vice versa.  In the panel (d) of Fig. 4 
we see the same behavior as in  the panel (c).  

\begin{figure}[tb]
\begin{center}
\includegraphics[width=55mm]{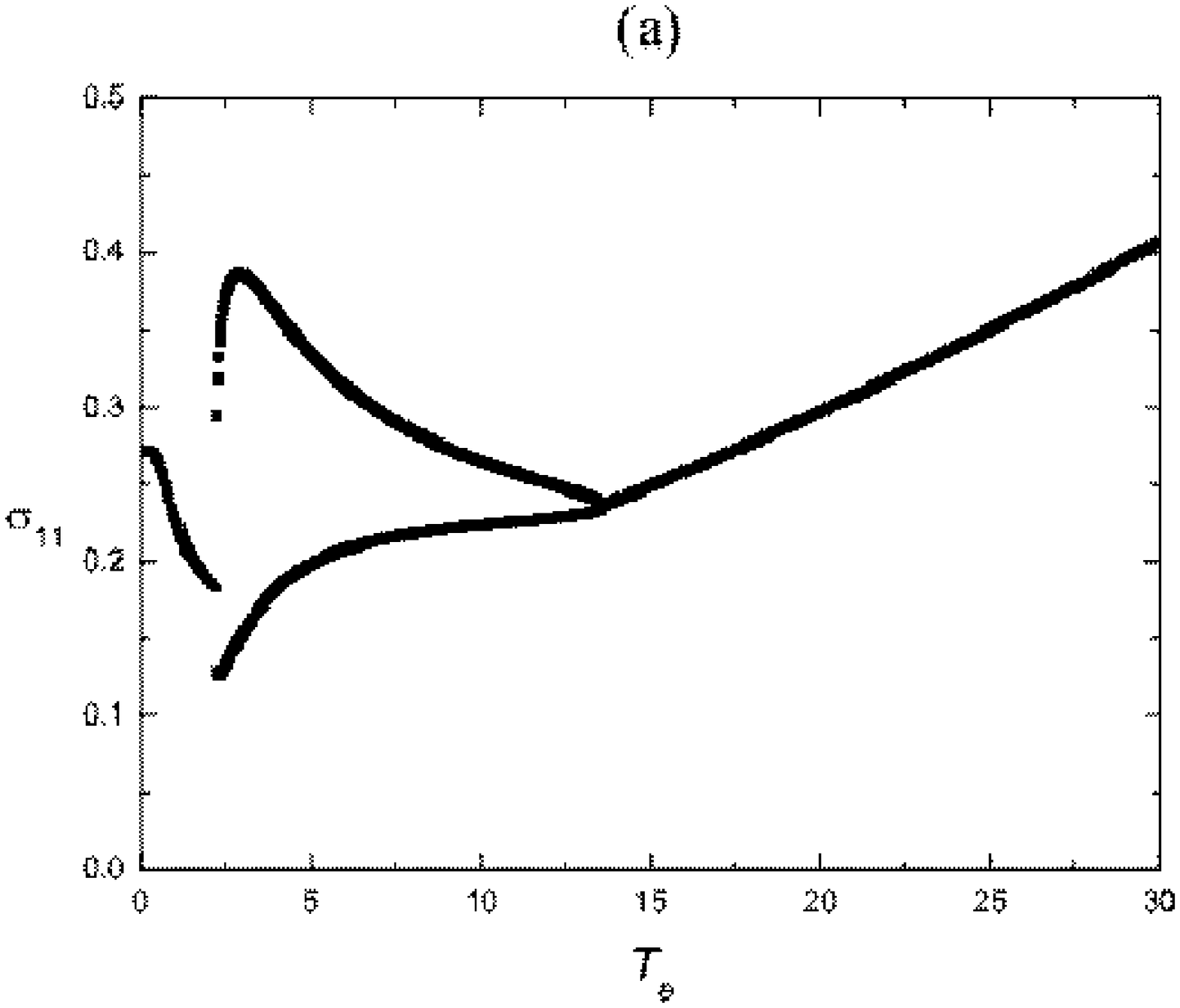}
\includegraphics[width=55mm]{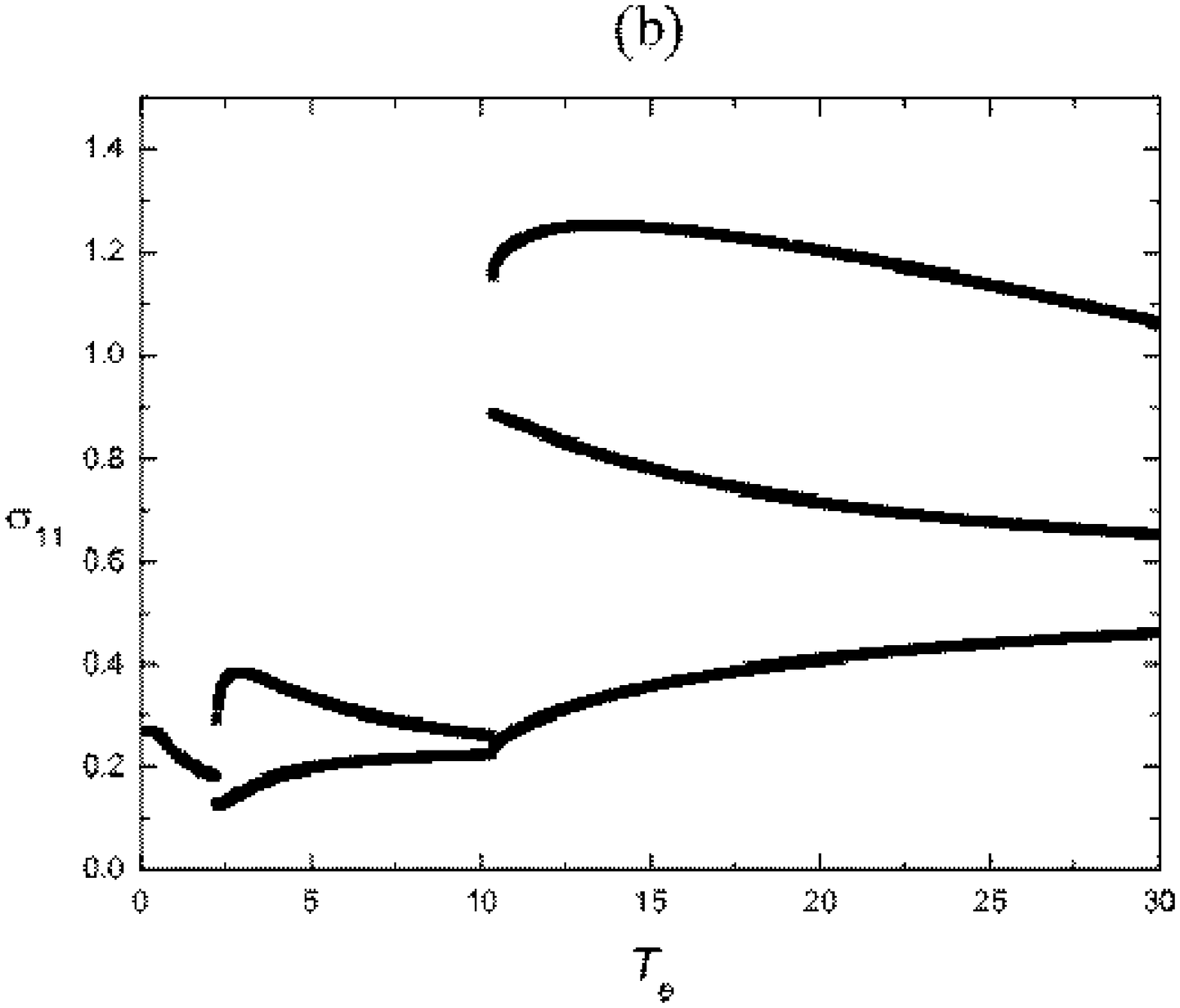}
\includegraphics[width=55mm]{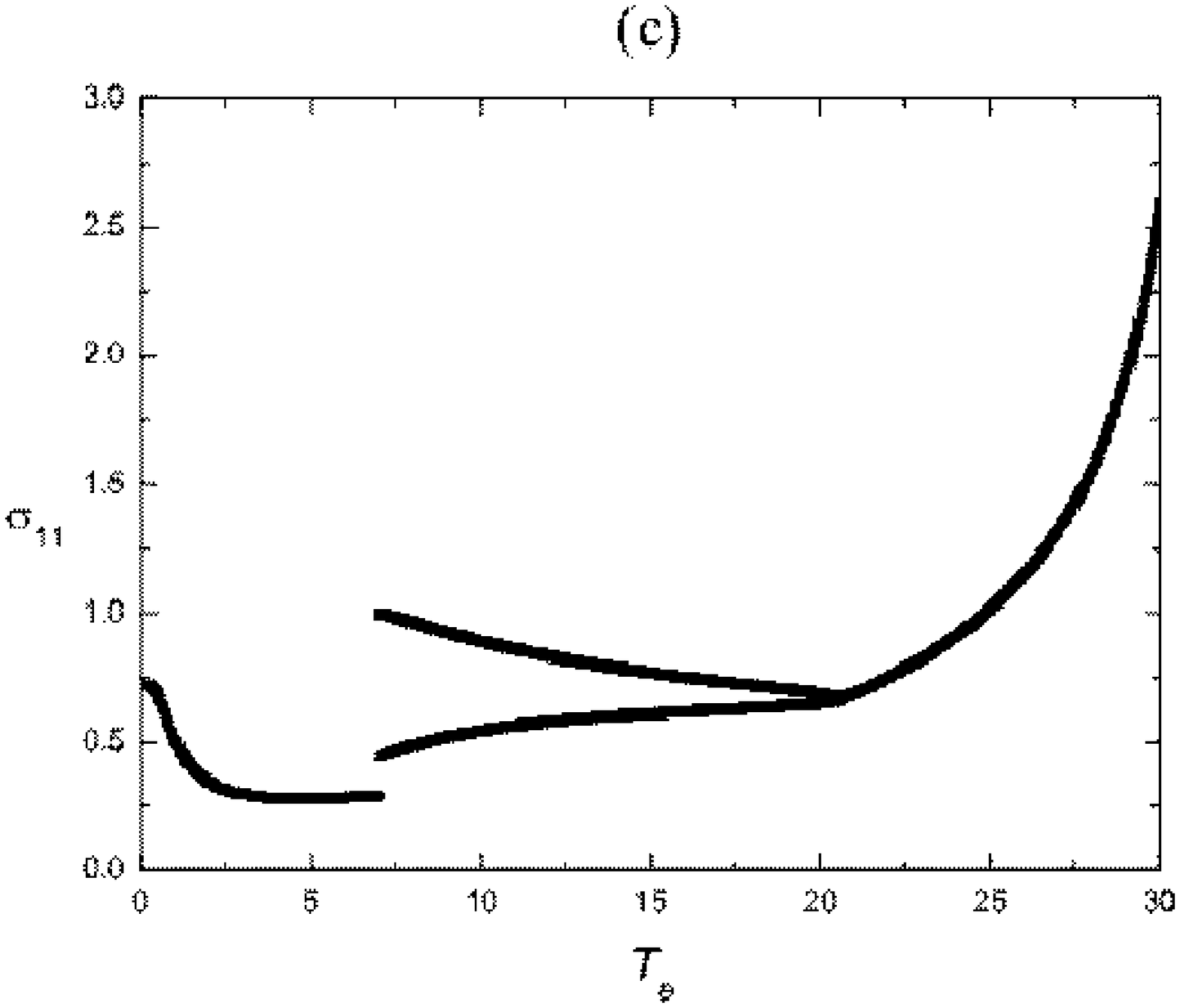}
\includegraphics[width=55mm]{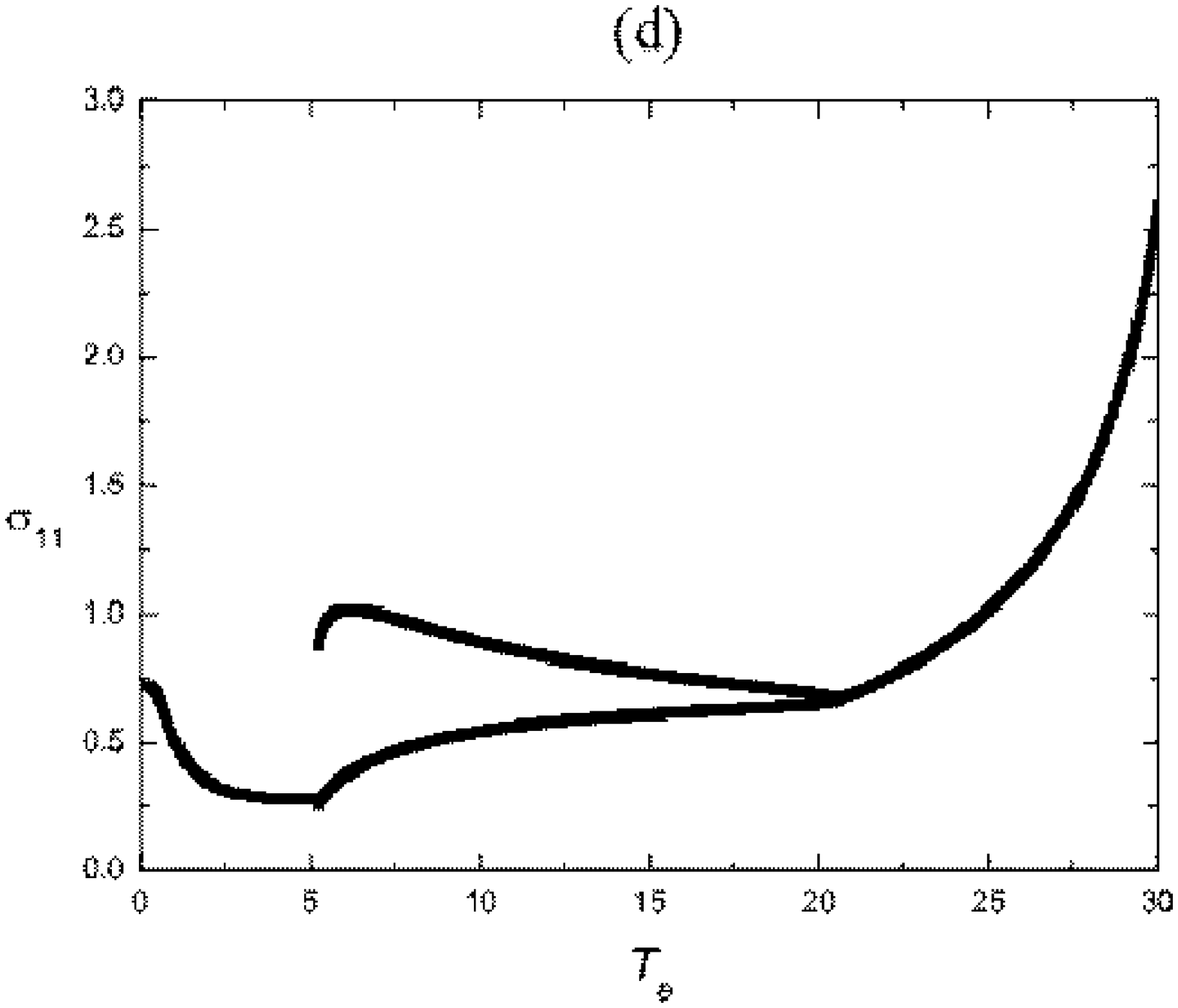}
\end{center}
 \caption{The bifurcations due to the constant and the composite
wakes for $\nu = 0.05$. In the panels (a) and (b) the equilibrium of
$\sigma_{11}$ for the constant wake with the parameters of $a = 10$ and 
$\sigma_{11}$ = $\sigma_{22}$ = 1 and $\sigma_{12}$ = 0 are shown,
respectively.   In the panels (c) and (d) the equilibrium of
$\sigma_{11}$ for the composite wake with the parameters of $a = 10$
and $\sigma_{11}$ = $\sigma_{22}$ = 1, $\sigma_{12}$ = 0 with $b$ =
5.0 are shown.  (a) and (c) represent
$\sigma_{11}$ when $T_e$ slowly changes from 0 to 30, and (b) and
(d) represent their reverse processes.} \label{fig5}
\end{figure}
Trackings of the mapping for the composite wake indicate that
when $b$ is small the system turns out to be similar to that
for the constant wake, while when the $b$ becomes larger, it
changes to another state.  Thus, in conclusion, when two
different kinds of wakes exist, the dynamical behavior of the
system can be changed, depending on the relative magnitude of the
strength of the wake: they may show different stability behaviors
in the parameter space from the case of a single wake.

\begin{figure}[tb]
\begin{center}
\includegraphics[width=55mm]{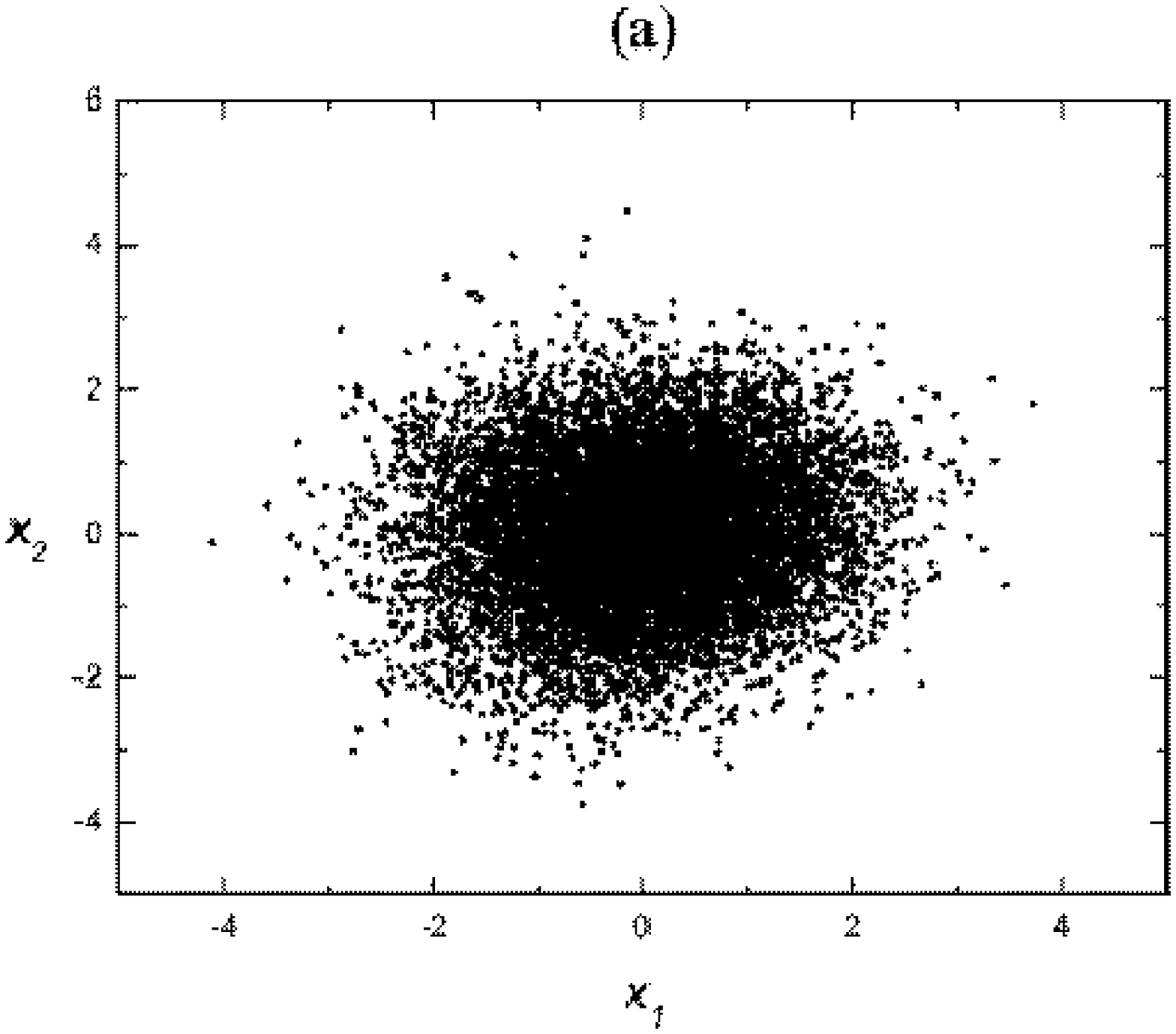}
\end{center}
 \caption{Phase-space distribution for the period-one
state from the multi-particle tracking for the linear wake 
function.  The parameters are given as $T_e = 10$, $\nu = 0.25$, $b =
0.25$ and $20,000$ turns.  The equilibrium state shows the same
distribution per 
turn.} \label{fig7} 
\end{figure}

\begin{figure}[tb]
\begin{center}
\includegraphics[width=55mm]{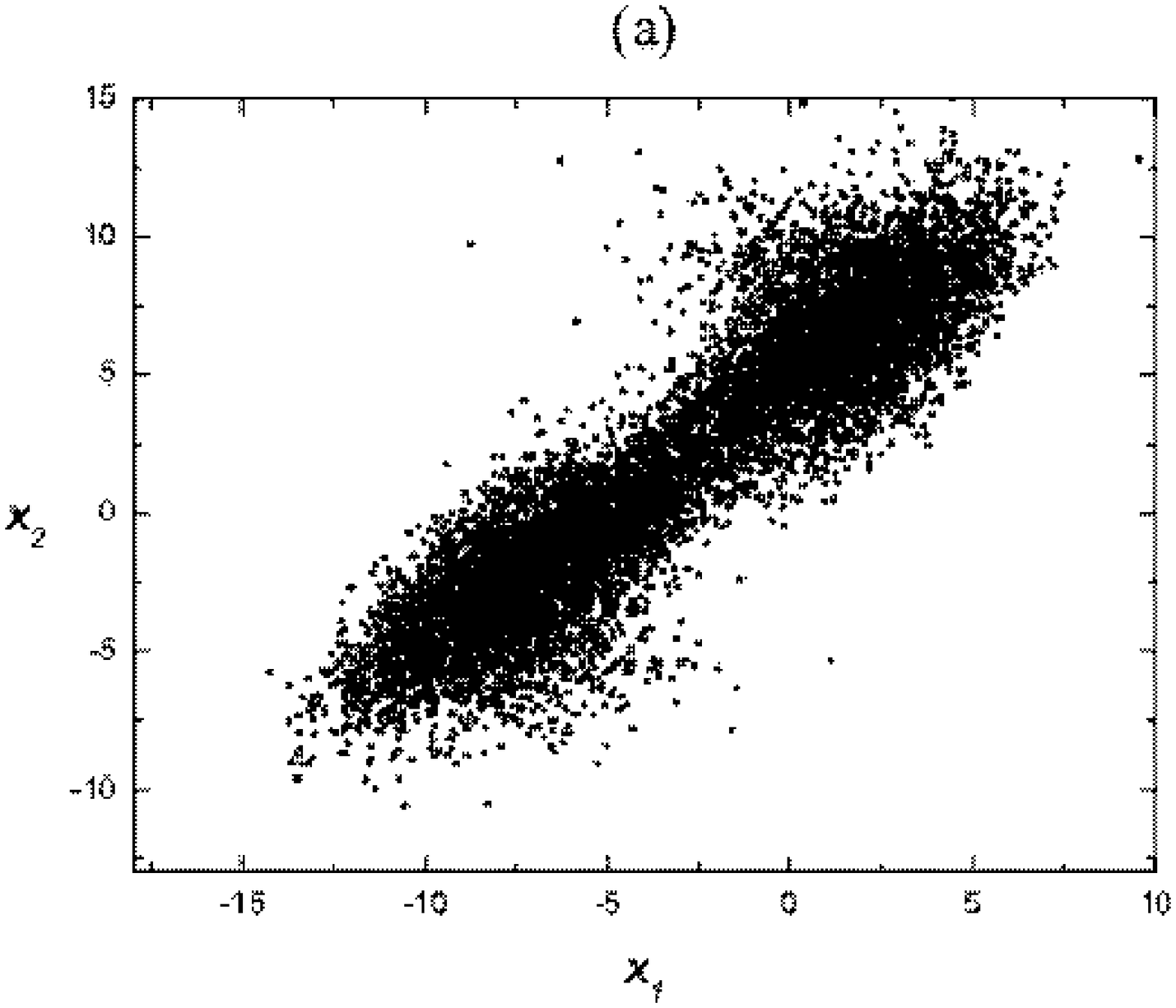}
\includegraphics[width=55mm]{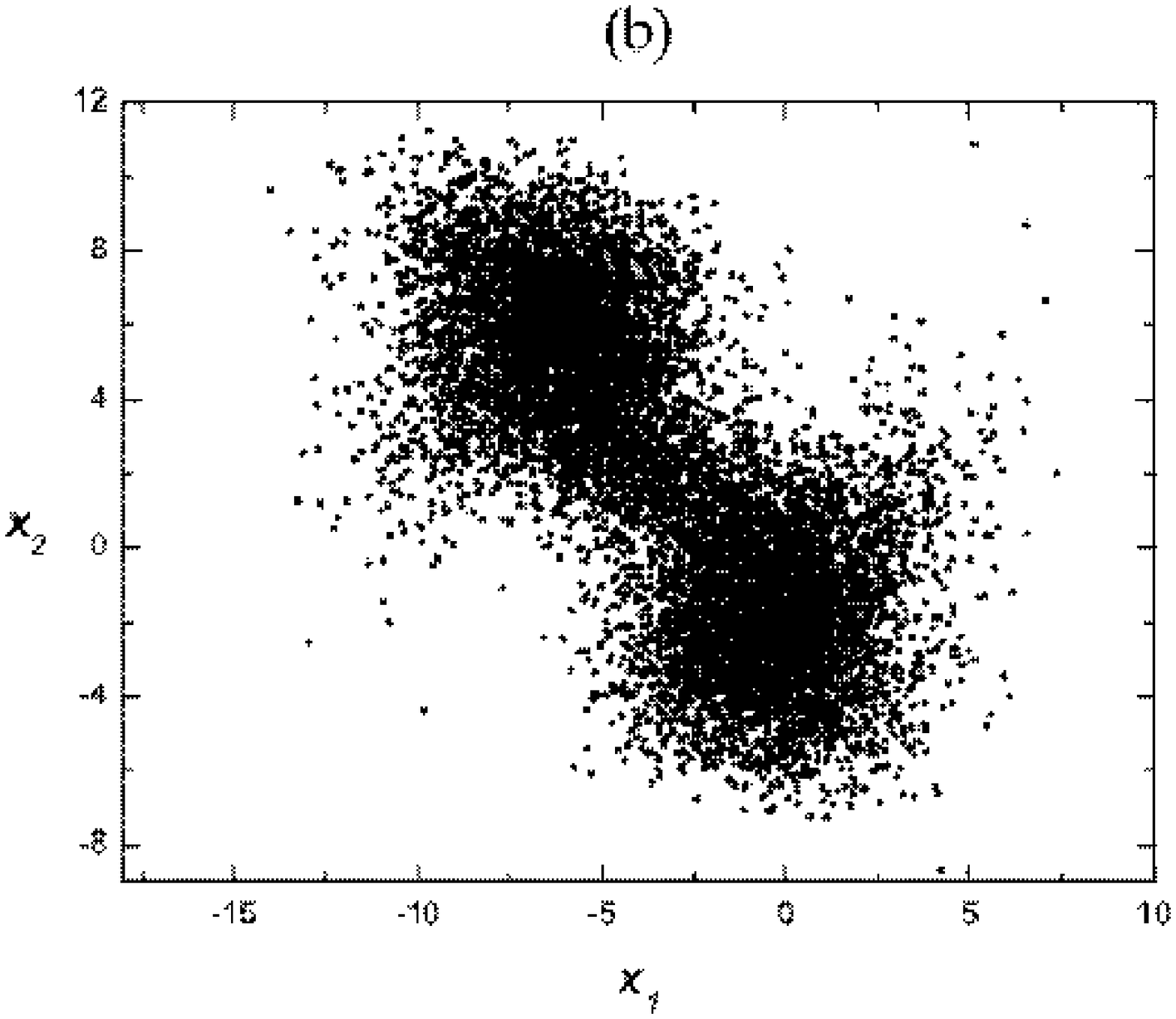}
\end{center}
\caption{Phase-space distribution for the period-two state from
the multi-particle tracking for the composite wake function.  The
parameters are given as $\nu$ = 0.20, $T_e$ = 22, $a$ = 8 and   $b$ = 
0.18.  (a) and (b) shows the phase space distributions at 19,999 
and  20,000 turns, respectively.  The equilibrium state shows the
same distribution per two turns.} \label{fig8}
\end{figure}

%%%%%%%%%%%%%%%%%%%%%%%%%%%%%%%%%
\section{Multi-particle Tracking}
%%%%%%%%%%%%%%%%%%%%%%%%%%%%%%%%%

In this section, we examine the reliability of the results from the
present model obtained in Secion 3.  The Gaussian Ansatz in the
model is considered as a particle distribution of a beam, however, the
real distribution can be rather far from the Gaussian one.  We
thus need to compare the Gaussian distribution with the multi-particle
tracking in order to see whether the results obtained from the model
are merely caused by the simplification of the model.  In the case of
the constant wake function, we use a sorting routine for the
calculation of the wake force acting on each particle, i.e.
the wake force $\phi$ that a given particle experiences is obtained by
counting the total number of particles preceding it.  As for
the linear wake function, we calculate individual distances of the
preceding particles in order to derive the wake force acting on
each particle.  Thus, the $\phi$ is obtained by summing the
distances of preceding particles.  The initial particle coordinates in 
the longitudinal phase space are generated as Gaussian random numbers
with zero mean and unit standard deviation.  We use Eq.~(5) to
perform the multi-particle tracking in the phase-space coordinates of
$10000$ macro-particles. 

First, we observe the equilibrium states of the particle
distribution in the case of the linear wake.  The phase space which
brings out the multi-particle tracking for $\nu=0.25$ is depicted 
in Fig.~5.  It is obtained by tracking 10000 particles after 20000
turns.  It is shown that the equilibrium state in the
multi-particle tracking accounts for the period-one state.  It is also 
examined that the only equilibrium state for the linear wake is
found in the period-1 state in the parameter of the synchrotron
tunes ranging from $0.01$ to $0.3$, irrespective of the initial
conditions.  It is in a good agreement with that of the Gaussian
model. 

We also find the equilibrium states of the particle distribution 
due to the composite effect of the constant and the linear
wakes.  In the panels (a) and (b) of Fig. 6, the phase space yielding
the results of the multi-particle tracking for $\nu=0.2$ is 
plotted.  It is obtained by tracking 10000 particles after 20000
turns. The panels (a) and (b) of Fig. 6 show the period-two state from
19999 turns and 20000 turns, respectively.  The equilibrium
distributions with the period-two state presents the same distribution
by two turns.  It is also shown that when $b$ becomes larger,
the equilibrium states in the multi-particle tracking produces the
transition from the period-two state to the period-one one and vice
versa.  Here, we see that periodic states shown by the
Gaussian model also appear in the multi-particle tracking.

As a result, we find that the results from the Gaussian model 
are in a qualitative agreement with those from the multi-particle 
tracking in the presence of the linear wake: The existence of the
period-doubling bifurcation and the transition between the
periodic states.  Note that in the present work we only have evaluated
the results of the multi-particle tracking for $\nu=0.2$ in order to
show that the simple Gaussian model with the composite wake explains
well the behaviors of periodic states and the transition   
between them, compared to the results from the multi-particle tracking 
for $0.01 \leq \nu \leq 0.3$.

%%%%%%%%%%%%%%%%%%%%%%%%%%%%%%%%%%%
\section{Discussion and Conclusion}
%%%%%%%%%%%%%%%%%%%%%%%%%%%%%%%%%%%

In the present work, we have investigated the nonlinear dynamical
behaviors and stable periodic states on the longitudinal beam
distribution, using both the Gaussian model and multi-particle
tracking.  It was shown that these periodic states have bifurcations to
other periodic states.  These behaviors are observed in both the
Gaussian model and multi-particle tracking.  The model calculations
showed the period-one state in the dynamic state of a particle distribution 
in a beam for the linear wake. It is also confirmed by the 
multi-particle tracking method.  When both the constant wake and linear
wake simultaneously exist, the parameter space in which the periodic
states exist can be changed by the magnitude of the linear wake.  It
was concluded that the Gaussian model seems to be useful in
investigating qualitatively the particle distribution in the
longitudinal phase space.  

One of the main purposes in the present work was to investigate the
effects of the composite cases taking into account the constant and
linear wakes.  We examined the dynamical behaviors of the equilibrium 
bunch length and their stability for a localized constant and linear
wakes.  Though the Gaussian model seems to be too simple, the results
from the model are in a good agreements with those from the
multi-particle tracking method.  When both a constant and  
linear wakes exist in a ring, it was shown that the transition 
between the periodic states depends on the relative strengths of
the two wakes. 

\section*{Acknowledgments}
The work of HChK was supported by Pusan National University.
The work of JKA was supported by Korea Institute of Science \& Technology
Evaluation and Planning(KISTEP) and Ministry of Science \& Technology
(MOST), Korean government through its BAERI program.
JKA is grateful to the PAL, where a part of the present work has been
achieved.

\begin{table} [p]

\caption{Synchrotron tune and stable periodic states for
   the case of the constant wake function.  }
\begin{center}
\begin{tabular} {ll}  \hline
  Synchrotron tune             & Existing periodic state            \\  \hline
  $0.244 \leq \nu \leq 0.30$      & 1                                  \\
    $0.153 \leq \nu \leq 0.243$     & 1;\ 2;\ 1-2                            \\
   $0.103\leq \nu \leq 0.152$       & 1;\ 2;\ 1-2;\ 1-3;\ 2-3                    \\
  $0.01 \leq \nu \leq 0.102$     & 1;\ 2;\ 1-3;\ 1-4;\ 2-3;\ 1-3-4;\ 2-3-4        \\ \hline
 \end{tabular}
\end{center}
\end{table}

\end{document}